\documentstyle[epsf]{l-school}
\newcommand{\vect}[1]{\mbox{\boldmath $#1$}}
\newcommand{\sol}{M_{\odot}}
\newcommand{\half}{\frac{1}{2}}
\newcommand{\rscript}[1]{\mbox{\scriptsize #1}}
\newcommand{\rhoh}{\rho_{_H}}
\newcommand{\rhob}{\rho_{_B}}
\newcommand{\rhon}{\rho_{_N}}
\newcommand{\gammaT}{\widetilde{\gamma}}
\newcommand{\gammaH}{\widehat{\gamma}}
\newcommand{\betaT}{\widetilde{\beta}}
\newcommand{\KT}{\widetilde{K}}
\newcommand{\JT}{\widetilde{J}}
\newcommand{\DT}{\widetilde{D}}
\newcommand{\LaplaceT}{\widetilde{\triangle}}
\newcommand{\auz}{\alpha u^0}
\newcommand{\sgamma}{\sqrt{\gamma}}
\newlength{\tempwidth}

\begin{document}
\markboth{K. Oohara \& T. Nakamura}{COALESCENCE OF BINARY NEUTRON STARS}
\title{Coalescence of Binary Neutron Stars}
\author{K. Oohara\inst{1} and T. Nakamura\inst{2}}
\institute{\inst{1} Department of Physics, \\
Niigata University, \\
Ikarashi, Niigata, 950--21, \\ Japan
\\
\inst{2} Yukawa Institute for Theoretical Physics, \\
Kyoto Univiersity, \\
Kyoto, 606--01, \\ Japan
}

\maketitle

\section{INTRODUCTION}
The most important sources for laser-interferometric
gravitational-wave detectors like LIGO or VIRGO are catastrophic
events such as coalescence of a neutron-star binary. A binary system
evolves quasi-stationarily and the wave pattern is regular up to the
onset of coalescence. In {\em the last three minutes} of coalescing
binaries, comparison of observed wave patterns with theoretical
templates will hence make it possible to determine mass and spin of
stars. In this phase, the post-Newtonian expansion will give good
templates since $v_{\hbox{\footnotesize rot}}/c$ is as small as 0.1.
In the final phase of the coalescence, however, there is no such small
parameters. Numerical relativity is one of methods for studying such
nonlinear phase of coalescing binary neutron stars; we call it {\em
  the last three milliseconds}. In order to know the characteristics
of the waves in the last three milliseconds, general relativistic (GR)
calculations are required. However, coalescence of binary stars is a
completely non-axisymmetric, 3 dimensional (in space) event and 3D GR
calculation requires great powers of computers.  We should hence make
progress step by step. First of all, we used a Newtonian hydrodynamics
code including radiation reaction of gravitational waves and then we
considered post-Newtonian correction, which is called (1+2.5)
post-Newtonian calculation, since a radiation reaction potential is
expressed as 2.5 post-Newtonian terms. These calculations give a lot
of perspectives on gravitational wave patterns and dynamics of
coalescing events. Finally we began to attack a 3D GR code for the
final phase of coalescing binary neutron stars.

In this lecture, we describe results of numerical simulations of
coalescing binary neutron stars using Newtonian and post-Newtonian
hydrodynamics code and then discuss recent development of our 3D GR
code for the last three milliseconds.

\section{POST-NEWTONIAN SIMULATIONS}

\subsection{(1+2.5) Post-Newtonian Hydrodynamics Equations}

The equations of motion including general relativistic effects up to
order $(v/c)^2$ is called the first post-Newtonian (1PN)
approximation.  However, gravitational damping effects start at order
$(v/c)^5$, second-and-a-half post-Newtonian (2.5PN) terms, and the
formulation for adding the effects to the equation of motion depends
on the gauge condition. For example, the radiation reaction is
represented by adding to the standard Newtonian potential a
``radiation-reaction'' potential proportional to the fifth time
derivatives of the quadrupole moments \cite{MTW73}.

In numerical calculations, however, it is hard to calculate the fifth
time derivatives with satisfactory accuracy.  Blanchet, Damour \&
Sch\"afer \cite{BDS90} presented (1+2.5)PN hydrodynamics equations
including radiation damping effects, where we need up to the third
time derivatives of the quad\-ru\-pole moments.  Following them, we
use the evolution equations given by
\begin{eqnarray}
  \partial_t \, \rho + \partial_j \left( \rho v^j \right) & = & 0,
   \label{eq:evol} \\
   \partial_t \left( \rho w_i \right) + \partial_j \left( \rho w_i v^j
   \right) & = & F_i^{\rscript{press}} + F_i^{\rscript{1PN}} +
   F_i^{\rscript{reac}},
   \label{eq:motion} \\
   \partial_t \left( \rho
     \varepsilon \right) + \partial_j \left( \rho \varepsilon v^j
   \right) & = & - p \, \partial_j v^j,
   \label{eq:enrg}
\end{eqnarray}
where $\rho$, $\varepsilon$ and $p$ are the coordinate rest-mass
density, the internal energy density and the pressure, respectively.
In order to express a hard equation of state for a neutron star, we
use a polytropic equation of motion
\begin{equation}
  \label{eq:eosg2}
  p = ( \gamma - 1) \rho \varepsilon
\end{equation}
with $\gamma = 2$.  The quantity $v^i$ denotes the 3-velocity and
$w_i$ is the ``momentum per unit rest-mass.'' In terms of $w_i$, the
3-velocity $v^i$ is given by
\begin{equation}
  v^i = \left( 1 - \frac{\beta}{c^2} \right) w_i + \frac{1}{c^2} A_i +
  \frac{4G}{5 c^5} w_j Q_{ij}^{[3]}, \label{eq:viwi}
\end{equation}
where
\begin{equation}
  Q_{ij}^{[3]} = \mbox{STF} \left\{ 2 \int \left[ p \partial_i w_j - 2
      \rho w^i \partial_j \psi + x^i \partial_j \psi \partial_k ( \rho
      w^k ) - \rho x^i \partial_j \psi_t \right] dV \right\},
\end{equation}
which is the third time derivative of the reduced quadrupole moment of
the system with the addition of corrections of order $O(c^{-2})$
\cite{Finn}. Here STF is the operator which takes the symmetric,
trace-free part of any two-index object $A^{ij}$:
\begin{equation}
 \mbox{STF} \left\{ A^{ij} \right\} = \frac{1}{2} A^{ij}
  + \frac{1}{2} A^{ji} - \frac{1}{3} \delta^{ij} A^{kk}.
\end{equation}
The forces are given by
\begin{eqnarray}
  & F_i^{\rscript{press}} = & - \partial_i \left[ \left(
      1+\frac{\alpha}{c^2} \right) p \right], \nonumber \\
  & F_i^{\rscript{1PN}} = & - \rho \left[ \left( 1+\frac{\delta}{c^2}
    \right) \partial_i \psi + \frac{1}{c^2} \, \partial_i \psi_2 +
    \frac{1}{c^2} \, w_j \partial_i A_j \right] ,
  \label{eq:force} \\ 
  & F_i^{\rscript{reac}} = & - \frac{1}{c^5} \, \rho \, \partial_i
  \psi_r.  \nonumber
\end{eqnarray}
Here $\psi$ is the Newtonian potential, $\alpha$, $\beta$, $\delta$,
$\psi_2$ and $A_i$ are 1PN quantities and $\psi_r$ is the
radiation-reaction potential (a 2.5PN quantity):
\vspace{\baselineskip}
\begin{eqnarray}
  \alpha & = & (3\gamma - 2) \psi - \frac{1}{2} \gamma w^2,
  \nonumber \\
  \beta & = & \frac{1}{2} w^2 + \gamma \varepsilon - 3 \psi,
  \nonumber \\
  \delta & = & \frac{3}{2} w^2 + (3\gamma - 2)
  \varepsilon + \psi, \\ A_i & = & U_i - \frac{1}{2} x^i \psi_t,
  \nonumber \\
  \psi_r & = & \frac{2}{5} G \left[ R - Q_{ij}^{[3]} x^i
  \partial_j \psi \right], \nonumber
\end{eqnarray}
where $w^2 = \delta^{ij} w_i w_j$ and $\psi_t$ is the time derivative
of $\psi$ with the addition of corrections of order $O(c^{-2})$.  To
compute $\psi$, $\psi_t$, $\psi_2$, $R$ and $U_i$, we must solve seven
Poisson equations at each time step:
\begin{eqnarray}
   & \Delta \psi & = 4 \pi G \rho,  \label{eq:np} \\
   & \Delta \psi_t & = - 4 \pi  G \partial_j \left( \rho w_j \right),
                                    \label{eq:npt} \\
   & \Delta \psi_2 & = 4 \pi G \rho \delta, \label{eq:psi2} \\
   & \Delta R & = 4 \pi G Q_{ij}^{[3]} x^i \partial_j \rho,
                                    \label{eq:rpot} \\
   & \Delta U_i & = - 4 \pi G \left( 4 \rho w_i
    + \frac{1}{2} \, x^i \partial_j ( \rho w_j ) \right). \label{eq:Ui}
\end{eqnarray}
The energy flux of the gravitational waves is given by
\begin{equation}
  \frac{dE}{dt} = -\, \frac{G}{5c^5} \, Q_{ij}^{[3]} \, \frac{d}{dt}
  I_{ij},
\end{equation}
where
\begin{equation}
 I_{ij} = \mbox{STF} \left\{ 2 \int \rho \left[ w_i w_j
  - x^i \partial_j \psi \right] dV \right\}, \label{eq:qp2t}
\end{equation}
which is the second time derivative of the reduced quadrupole moment
with the addition of corrections of order $O(c^{-2})$.  The standard
quadrupole formula gives the amplitude $h^{\rscript{TT}}_{ij}$ of the
gravitational waves in the transverse-traceless gauge \cite{MTW73}
\begin{equation}
 h^{\rscript{TT}}_{ij} = \frac{2}{r} \left( P_{im} P_{jn} -
   \frac{1}{2} P_{ij} P_{mn} \right) \frac{d^2 Q_{mn}}{d t^2} ,
 \label{eq:gramp}
\end{equation}
where $Q_{mn}$ is the reduced mass quadrupole moment
\begin{equation}
 Q_{mn} = \mbox{STF} \left\{ \int \rho x^m x^n \, dV \right\}
\end{equation}
and $P_{ij} = \delta_{ij} - n_i n_j$ is the projection operator onto
the plane transverse to the outgoing wave direction, $n_i = x^i / r$.
For consistency with post-Newtonian accuracy, we must take into
account the relativistic corrections up to the relative order
$(v/c)^2$.  This requires the time derivatives of the mass octupole,
mass $2^4$-pole, current quadrupole and current octupole \cite{BDS90}.
However we neglect these corrections since the gravitational wave
amplitude is evaluated with quantities at each time step and small
truncation errors above `Newtonian' accuracy do not accumulate.  With
this accuracy, we can use $I_{ij}$ defined by Eq.(\ref{eq:qp2t}) in
place of $\ddot{Q}_{ij}$.  From Eq.(\ref{eq:gramp}), two polarizations
are given by
\begin{eqnarray}
  h_{+} & = & \frac{1}{r} \left( I_{\hat{\theta} \hat{\theta}}
 - I_{\hat{\phi} \hat{\phi}} \right) , \nonumber \\
  h_{\times} & = & \frac{2}{r} \, I_{\hat{\theta} \hat{\phi}} ,
\end{eqnarray}
where $I_{\hat{i} \hat{j}}$ is the quantity in the orthonormal basis
\begin{eqnarray}
  I_{\hat{\theta} \hat{\theta}} & = & \left( I_{xx} \cos^2 \phi
   + I_{yy} \sin^2 \phi + 2 I_{xy} \sin \phi \cos \phi \right) \cos^2
 \theta \nonumber \\
  & & + I_{zz} \sin^2 \theta - 2 \left( I_{xz} \cos \phi + I_{yz} \sin
    \phi \right) \sin \theta \cos \theta  , \nonumber \\[.25em]
  I_{\hat{\phi} \hat{\phi}} & = & I_{xx} \sin^2 \phi + I_{yy} \cos^2
  \phi - 2 I_{xy} \sin \phi \cos \phi , \\[.25em]
  I_{\hat{\theta} \hat{\phi}} & = & \left( I_{yy} - I_{xx} \right)
   \cos \theta    \sin \phi \cos \phi
   + I_{xy} \cos \theta ( \cos^2 \phi - \sin^2 \phi )   \nonumber \\
  & & + I_{xz} \sin \theta \sin \phi - I_{yz} \sin \theta \cos \phi
  \nonumber
\end{eqnarray}

\subsection{Numerical Methods}

The equations are discretized by the finite difference method (FDM)
with a uniform Cartesian grid. The evolution equations are integrated
using van Leer's scheme \cite{vanLeer} with second-order accuracy in
space.  In order to make the scheme stable, a monotonicity condition,
the so-called TVD (Total Variation Diminishing) limiter \cite{Harten},
is imposed on this scheme \cite{ON92}. In order to achieve
second-order accuracy in time we adopt a two-step procedure as
follows. To describe our code, we write
Eqs.(\ref{eq:evol})--(\ref{eq:enrg}) as
\begin{equation}
  \partial_t Q + \partial_j ( Q v^j ) = f(Q),
\end{equation}
where $Q$ is a 5-dimensional vector defined by
\begin{equation}
  Q = ( \rho, \rho w_i, \rho \varepsilon ).
\end{equation}
The elements of 5-vector $f(Q)$ are given by the right hand sides of
Eqs.(\ref{eq:evol})--(\ref{eq:enrg}). In order to obtain $Q^{n+1}
\equiv Q(t^n + \Delta t^n)$, first we calculate the quantity $Q^{n +
  \half} \equiv Q(t^n + \Delta t^n/2)$ from $Q^n$ as
\begin{equation}
  Q^{n + \half}  =  Q^n + F(Q^n) \, \frac{\Delta t^n}{2} ,
\end{equation}
and then calculate $Q^{n+1}$ from $Q^n$ and $Q^{n + \half}$ as
\begin{equation}
  Q^{n + 1}  =  Q^n + F(Q^{n+ \half}) \,  \Delta t^n ,
\end{equation}
where $F(Q)$ is the flux given by $f(Q)$.

In order to treat shock waves, we use a tensor artificial viscosity
given by
\begin{equation}
 p_{ij} = \left\{
 \begin{array}{ll}
   \! \rho \, \ell^2 ( \partial_k v^k )
       \mbox{STF} \left\{ 2 \, \partial_i v^j \right\}
     & \mbox{if } \partial_k v^k < 0, \\[.5em]
   0 & \mbox{otherwise},
 \end{array} \right.
\end{equation}
where $\ell$ is an appropriate number with units of length.  The gas
pressure $p$ is replaced by $P_{ij} = p \delta_{ij} + p_{ij}$.  The
Poisson equations are solved using the MICCG (Modified Incomplete
Cholesky-decomposition and Conjugate Gradient) method
\cite{MV77,Gust,Ushiro,Vorst,MOK85}, which is fully vectorized using
the hyperplane method proposed by Ushiro \cite{MOK85}.

We have performed various tests for this code \cite{ON89, NO89b},
which include; (1) free transportation of a dust cube of a homogeneous
or Gaussian density distribution, (2) a 1D Riemann shock tube, (3) a
point explosion in the air, (4) local conservation of specific angular
momentum for an axially symmetric collapse and (5) collapse of a
homogeneous dust ellipsoid.  Our code passed these tests with
sufficiently good accuracy.

\subsection{Newtonian Calculation}

First we performed an extensive series of numerical calculations using
Newtonian hydrodynamics including the radiation damping effects, where
terms of order $O(c^{-2})$ in Eqs.(\ref{eq:viwi})--(\ref{eq:force}) are
neglected while terms of order $O(c^{-5})$ are kept to include the
radiation damping effects.

\subsubsection{Coalescence of Neutron Stars without Spin}

We first consider initial data comprising of two neutron stars
rotating rigidly around the center of mass. The binary system is
assumed to be in rotational equilibrium at the initial time.

In order to obtain hydrostatic equilibrium models of close neutron
star binaries, we set $\partial_t = 0$, $F_i^{\rscript{reac}} = 0$ and
$v^i = w_i$ in the Newtonian version of
Eqs.(\ref{eq:evol})--(\ref{eq:enrg}).  Here the gravitational radiation
damping effects are neglected since a binary system evolves
quasi-stationarily up to the onset of coalescence.  Assuming the
pressure $p$ is given by $p = K \rho^{\gamma}$ and the two stars are
in a synchronized circular orbit around the center of mass $(x_0, y_0,
0)$, that is, the velocity is given by $v^i = ( -(y-y_0) \Omega,
(x-x_0) \Omega, 0)$, we have the equation that equilibrium models
should satisfy: 
\bigskip
\begin{equation}
 \nabla \left[ \psi + H - \frac{1}{2}
   \left\{ (x-x_0)^2 + (y-y_0)^2 \right\} \Omega^2
        \right] = 0, \label{eq:equil}
\end{equation}
\vskip 0.75em \noindent
where $H$ is the enthalpy,
\vskip 0.5em
\begin{equation}
   H = \frac{\gamma}{\gamma - 1} K \rho^{\gamma - 1} \  .
\end{equation}
\vskip 1em \noindent
A solution of Eq.(\ref{eq:equil}) for each star is given by
\begin{equation}
  \psi + H - \frac{1}{2} \left\{ (x-x_0)^2 + (y-y_0)^2 \right\} \Omega^2
  = C_i \mbox{\hspace{2ex} (for $i=$1, 2)} \label{eq:equil1}
\end{equation}
where $C_1$ and $C_2$ are different constants in general.  Since the
position of the center of mass can be set freely, we set $y_0
=0$. Instead of setting $x_0$, however, we fix the positions $x^s_1$
and $x^s_2$ where the surface of each star intersects the $x$-axis
between the stars; $\rho(x^s_1,0,0) = \rho(x^s_2,0,0) = 0$. In
addition, we fix the center of each star $x^c_1$ and $x^c_2$ and the
density there; $\rho(x^c_1,0,0) = \rho^c_1$ and $\rho(x^c_2,0,0) =
\rho^c_2$.  A self-consistent solution is determined by an iterative
method.  First, setting the constants $\gamma$ and $K$ in the equation
of state, we give an initial guess of the density distribution,
usually as two spherical polytropes. Then we repeat the following
steps until convergence:
\bigskip

{\def\labelenumi{(\arabic{enumi})}
\begin{enumerate}
\item The potential $\psi$ is calculated as the solution of
Eq.(\ref{eq:np}).
\item $C_1$, $C_2$, $x_0$ and $\Omega$ are determined from
\settowidth{\tempwidth}{ $\displaystyle + H^c_1 $}
\begin{eqnarray}
  \psi^c_1 + H^c_1 - \, \frac{1}{2} \, (x^c_1 - x_0)^2 \Omega^2 & = &
  C_1 , \nonumber \\
  \psi^s_1 \mbox{\hspace{\tempwidth}} - \,
  \frac{1}{2} \, (x^s_1 - x_0)^2 \Omega^2 & = & C_1 , \nonumber \\ 
  \psi^c_2 + H^c_2 - \, \frac{1}{2} \, (x^c_2 - x_0)^2 \Omega^2 & = &
  C_2 , \label{eq:equilibrium} \\
  \psi^s_2 \mbox{\hspace{\tempwidth}} - \, \frac{1}{2} \,
  (x^s_2 - x_0)^2 \Omega^2 & = & C_2 ; \nonumber
\end{eqnarray}
or explicitly
\begin{eqnarray}
  x_0 & = & \frac{A_2 \{ (x^c_1)^2-(x^s_1)^2 \} - A_1 \{
    (x^c_2)^2-x(^s_2)^2 \} } {2 \{ A_2 (x^c_1 - x^s_1) -A_1 (x^c_2 -
    x^s_2) \} } \ , \nonumber \\ \Omega^2 & = & \frac{A_1}{(x^c_1 -
    x^s_1)(x^c_1+x^s_1- 2 x_0)} \ , \\ C_i & = & \psi^s_i -
  \frac{1}{2} \, (x^s_i - x_0)^2 \Omega^2 , \nonumber
\end{eqnarray}
where $\psi_{a} \equiv \psi(x_a, 0, 0)$, $H_a \equiv H(x_a,0,0)$ and
$A_i \equiv \psi^c_i - \psi^s_i + H^c_i$.
\item Using these values, we determine a new density distribution from
\begin{equation}
  H = \frac{\gamma}{\gamma - 1} K \rho^{\gamma - 1} = C_i - \psi + \,
  \frac{1}{2} \, \{ (x - x_0)^2 + y^2 \} \Omega^2 .
\end{equation}
Here we use the constant $C_1$ for $x > x_0$ and $C_2$ for $x < x_0$.
\end{enumerate}
}
\medskip

To be more realistic, we consider an infalling velocity due to the
gravitational radiation.  Assuming that the two stars are point masses
with separation $\ell$ in a circular orbit, $\ell$ decreases at a rate
\begin{equation}
  \partial_t \ell = - \, \frac{64 m_1 m_2 (m_1 + m_2)}{5 \ell^3}.
\end{equation}
We therefore add the infalling velocity given by this equation to the
initial stars.

We typically use a $141 \times 141 \times 131$ grid.  We assume
reflection symmetry with respect to the $z=0$ plane and consider the
region of $z \ge 0$ only, since the system becomes rather flat and
therefore needs a finer grid in the $z$-direction.  Calculations were
performed on a HITAC S820/80 supercomputer at the National Laboratory
for High Energy Physics (KEK).  Each calculation requires
approximately 900Mbytes of memory.
A typical CPU
time required is 100--200 hours for up to 50,000--90,000 time steps,
which corresponds to an event lasting 4--8 milliseconds.

\begin{figure}[tbp]
\begin{center}
  \leavevmode
  \epsfxsize=\textwidth \epsfbox{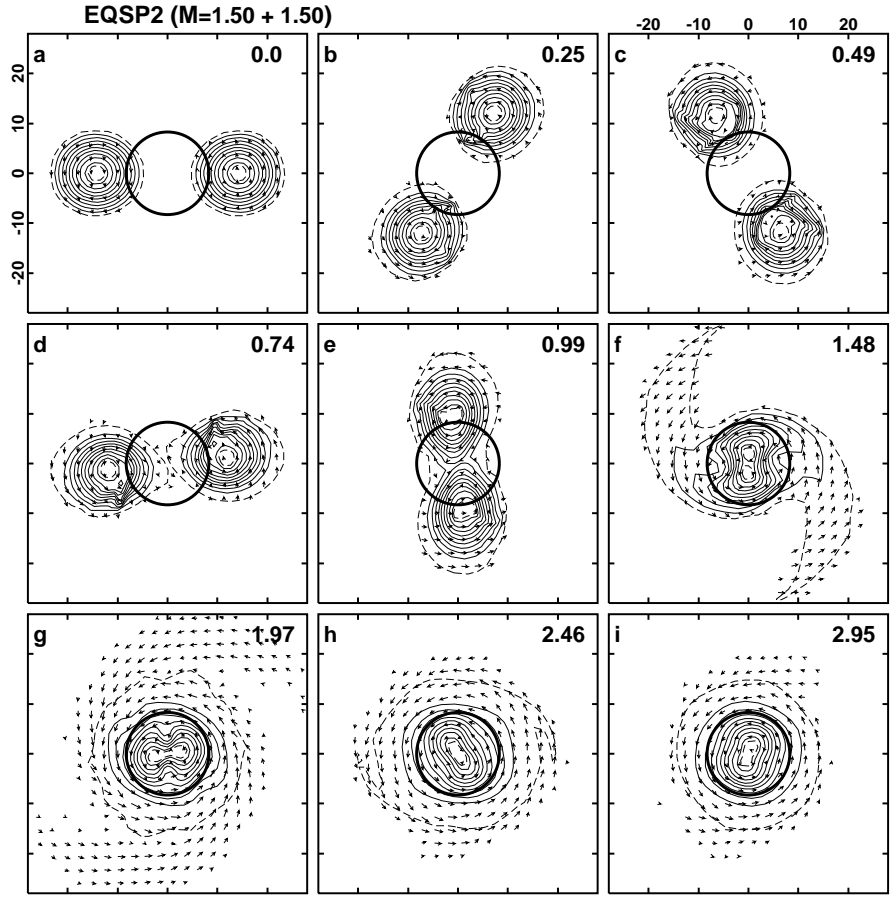}
  \caption{Density and velocity on the $x$-$y$ plane for EQSP2. The time
    in units of milliseconds is shown. Solid lines are drawn in steps
    of a tenth of the maximum density and the inner and the outer
    dashed lines indicate 19/20 and 1/20 of maximum density. Arrows
    indicate the velocity vectors of the matter. A fat line shows a
    circle of radius $2G M_t /c^2$ to show the size of a spherical
    black hole for comparison.} \label{fig:eqsp2-density}
\end{center}
\end{figure}

\begin{figure}[tbp]
\begin{center}
  \leavevmode
  \epsfxsize=.7\textwidth \epsfbox{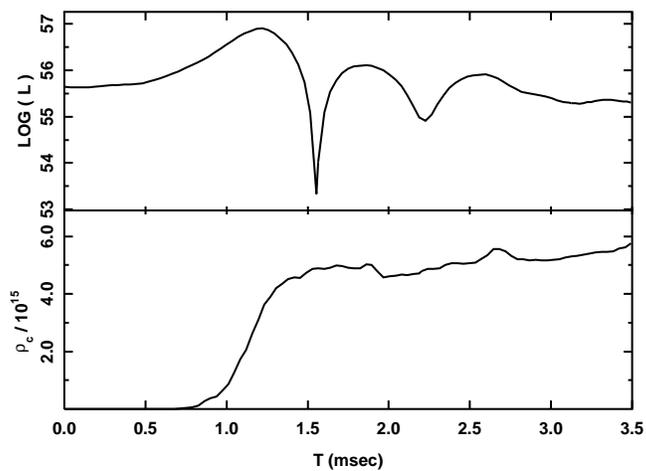}
  \caption{Luminosity (in units of erg/sec) and central density (in
    units of g/cm$^3$) as functions of time (in units of millisecond)
    for EQSP2.} \label{fig:eqsp2-flux}
\end{center}
\end{figure}

Here we show the results for neutron stars of $1.5\sol + 1.5\sol$
(EQSP2). \Video{EQSP2} Figure \ref{fig:eqsp2-density} shows the
evolution of the density and the velocity on the $x$-$y$ plane, where,
at the initial time, the radius of each star is 8.8km, the separation
between stars $\ell_0$ is 27km, the angular velocity $\Omega$ is $4.1
\times 10^3 \mbox{sec}^{-1}$.  In each figure only the central part is
represented, while the computational grid covers $[ -47 \mbox{km}, 47
\mbox{km}]$ in the $x$ and $y$ directions and $[0 \mbox{km}, 49
\mbox{km}]$ in the $z$ direction. Figure \ref{fig:eqsp2-flux} shows
the emitted energy $L$ and the central density $\rho_c$ as functions
of time.  The gravitational radiation takes the angular momentum away
from the system; in fact, $a/m \equiv J_t/(G M^2/c^3) = 0.64$ at first
and $a/m = 0.38$ finally.  Since the greatest part of the matter is
within the Schwarzschild radius $r = G M_t / c^2$ as shown in
Fig.\ref{fig:eqsp2-density}i, the final destiny of the system must be
a slowly rotating black hole. The energy emitted in gravitational
radiation is $1.6 \times 10^{53}$erg, which is 3\% of the total rest
mass.  This means that the efficiency of gravitational radiation is 30
times larger than the results for axisymmetric simulations by Smarr
\cite{Smarr} or Stark and Piran \cite{SP85}.

\begin{figure}[tbp]
\begin{center}
  \leavevmode
  \epsfxsize=.7\textwidth \epsfbox{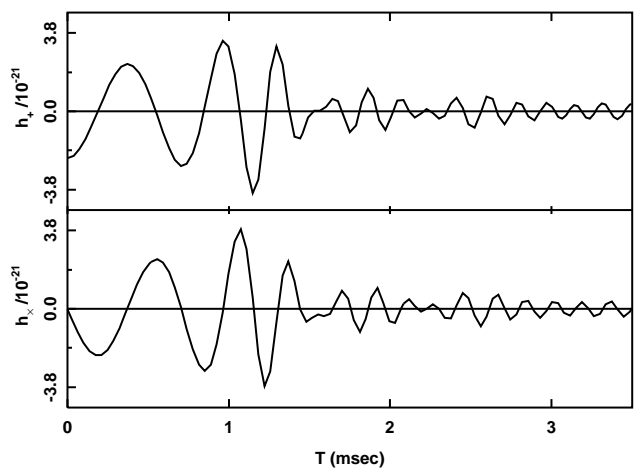}
  \caption{Wave forms of $h_{+}$ and $h_{\times}$ observed on the
    $z$-axis at 10Mpc for EQSP2.} \label{fig:eqsp2-wave}
\end{center}
\end{figure}
\begin{figure}[tbp]
\begin{center}
  \leavevmode
  \epsfxsize=\textwidth \epsfbox{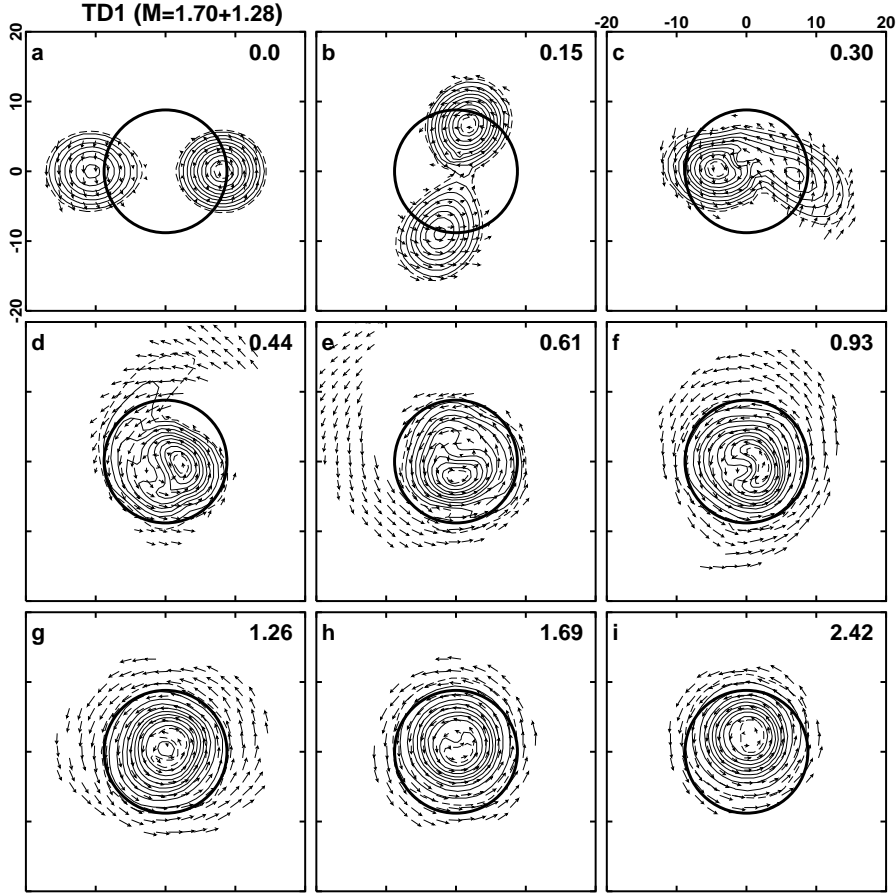}
  \caption{Density and velocity on the $x$-$y$ plane for
    TD1.}  \label{fig:td1-density}
\end{center}
\end{figure}

Figure \ref{fig:eqsp2-wave} shows the wave forms $h_{+}$ and
$h_{\times}$ observed along the $z$-axis, which are given by
\begin{eqnarray}
 h_{+} & = & \frac{1}{r} \left( I_{xx} - I_{yy} \right) ,
           \nonumber \\[.25em]
 h_{\times} & = & \frac{2}{r} I_{xy},
\end{eqnarray}
where $r$ is the distance between the event and the observer.
If the event occurs at 10Mpc from the earth, the maximum amplitude of
$h$ is $4.0 \times 10^{-21}$.

\begin{figure}[tbp]
\begin{center}
  \leavevmode
  \epsfxsize=.7\textwidth \epsfbox{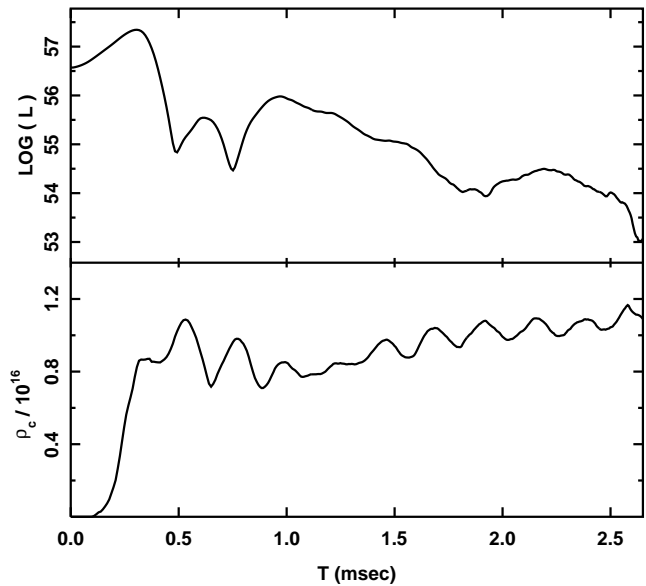}
  \caption{Luminosity (in units of erg/sec) and central density (in
    units of g/cm$^3$) as functions of time (in units of millisecond)
    for TD1.} \label{fig:td1-flux}
\end{center}
\end{figure}

Figures \ref{fig:td1-density} and \ref{fig:td1-flux} is the same as
Figs.\ref{fig:eqsp2-density} and \ref{fig:eqsp2-flux}, but for the
model TD1, where the masses of stars are $1.70\sol$ and $1.28\sol$,
the radii are 7km and 8km, $\Omega = 7.9 \times 10^3 \mbox{sec}^{-1}$
and $J_t = 5.1 G\sol^{2}/c$. \Video{TD1}

\subsubsection{Coalescence of Neutron Stars with Spin}

\begin{figure}[tbp]
  \begin{center}
    \leavevmode
      \epsfxsize=.4\textwidth \epsfbox{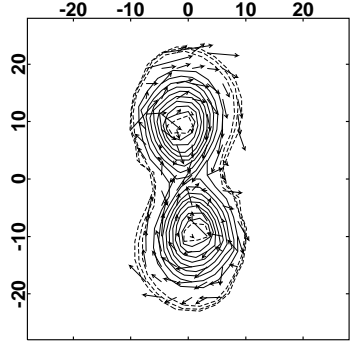}
    \caption{The same as
      Fig.\protect\ref{fig:eqsp2-density}e. The arrows show the
      velocity relative to the center of each star, which shows the
      spin retrograde to the orbital motion.}
    \label{fig:eqsp2-denr}
  \end{center}
\end{figure}

In the previous subsection, we place two neutron stars in a rotational
equilibrium state with rigid rotation around the center of mass as the
initial condition. However, in the absence of the viscosity, the
circulation $\Gamma$ of the system should be conserved:
\begin{equation}
  \label{eq:circulation}
  \Gamma = \int_S (\nabla \times \vect{v}) \cdot d \vect{S}
  = \oint_{\partial S} \vect{v} \cdot d \vect{\ell}
  = \mbox{const}.
\end{equation}
If the stars are assumed to rotate rigidly around the center of mass
at first, the circulation in the $z =$ constant plane is $2\Omega_0
S$, where $\Omega_0$ and $S$ are the initial angular velocity and the
area of the cross section of the stars, respectively. While two stars
approach each other, the area of the cross section of the stars does
not change so much, but the angular velocity $\Omega$ becomes much
larger than $\Omega_0$ as $\Omega \propto (\mbox{separation})^{3/2}$.
Then, the stars must have spin with the angular velocity $- \Omega$ to
conserve the circulation. This is illustrated in
Fig.\ref{fig:eqsp2-denr}. It is the same figure as
Fig.\ref{fig:eqsp2-density}e, but the arrows show the velocity
relative to the center of each star. 

To take into account this effect, we consider two neutron stars
rotating around the center of mass but also having spin angular
velocity retrograde to the orbital motion \cite{SNO92}.  In this case,
we don't know how to determine an equilibrium model because this is a
similar problem to determine the Dedekind configuration \cite{IM81}.
As for an axisymmetric rotating star, however, we can obtain the
equilibrium configurations. We here consider axisymmetric rotation
stars as the initial condition for each spinning star.  The
equilibrium configuration for a spinning star is determined by
Eq.(\ref{eq:equilibrium}) with $\psi_i$ and $\Omega$ being the
gravitational potential of each star alone and the spin angular
momentum, respectively, as well as $x_0 = x_i^c$.  Then we assume that
the orbital velocity of each star is the Keplarian one, $\Omega_K =
\sqrt{G (m_1 + m_2)/\ell_0^3}$, where $m_1$, $m_2$ and $\ell_0$ are
mass of each star and the separation of each star, and that $\Omega =
- \Omega_K$. This initial condition is consistent if the tidal force
is much smaller than the self-gravity.

Initially the orbital velocity $\hbox{\boldmath{$v$}}_K$ is given by
\begin{equation}
  v_K^x = - y \Omega_K, \ \ \ v_K^y = x \Omega_K
\end{equation}
and the spin velocity $\hbox{\boldmath{$v$}}_s$ is given by
\begin{eqnarray}
  v_s^x & = & y \Omega_K \nonumber \\[.5em]
  v_s^y & = & \!\!\! \left\{
    \begin{array}[c]{ll}
      {\displaystyle - \left( x - \frac{\ell_0}{2} \right) \Omega_K}
      & (x > 0) \\[1.5em]
      {\displaystyle - \left( x + \frac{\ell_0}{2} \right) \Omega_K}
      & (x < 0).
    \end{array}
    \right.
\end{eqnarray}
\begin{figure}[tbp]
\begin{center}
  \leavevmode
  \epsfxsize=\textwidth \epsfbox{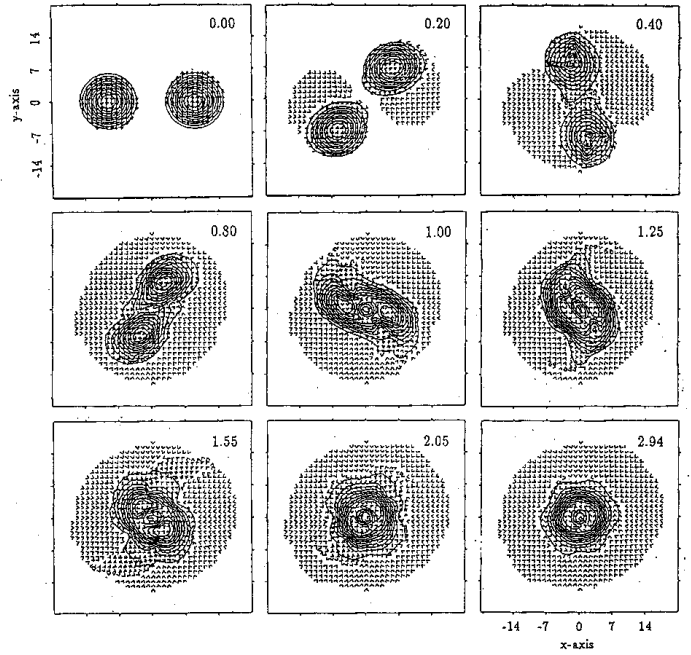}
  \caption{Density and velocity on the $x$-$y$ plane for SPIN1.}
  \label{fig:spin1-density}
\end{center}
\end{figure}%
Since $v^{x}_K + v^x_s = 0$ in the inertial frame, the velocity vector
points to only $y$-direction.  Figure \ref{fig:spin1-density} shows
the evolution of the density and the velocity on the $x$-$y$ plane,
where, at the initial time, the mass and the radius of each star are
$1.40\sol$ and 9km, the separation between stars $\ell_0$ is 27km.
The evolution sequence is almost the same as EQSP2 up to the onset of
coalescence.  After coalescence starts, however, the evolution of the
system is quite different; there is no spiral arm in the outer region
and the configuration becomes a nearly axially symmetric disk
soon. This is because the velocity in the outer region is smaller on
account of the spin retrograde to the orbital motion and the
centrifugal force is weaker than EQSP2. The stars are gradually
coalescing because the centrifugal force is enhanced in the inner
region, and the double core structure is kept for a long time. The
radiation reaction makes the double cores merge at last.

\subsection{Post-Newtonian Calculation}

Now we show results of a (1+2.5) post-Newtonian calculation we
performed in order to study general relativistic effects. We performed
also a Newtonian calculation with the same initial data for
comparison. The initial data includes two spherical neutron stars,
namely two relativistic polytropes with $\gamma=2$ whose surfaces just
touch each other. The mass and the radius of each star is $0.62\sol$
and 15km, respectively. The two stars are rotating around the center
of mass with angular velocity $\Omega=2.0 \times 10^3
\mbox{sec}^{-1}$. The total angular momentum, $J_t$, of the system is
$1.6G\sol^2/c$.
\begin{figure}[p]
  \begin{center}
    \leavevmode
    \leavevmode
    \epsfysize=.9\textheight \addtolength{\epsfysize}{-4em}
    \epsfbox{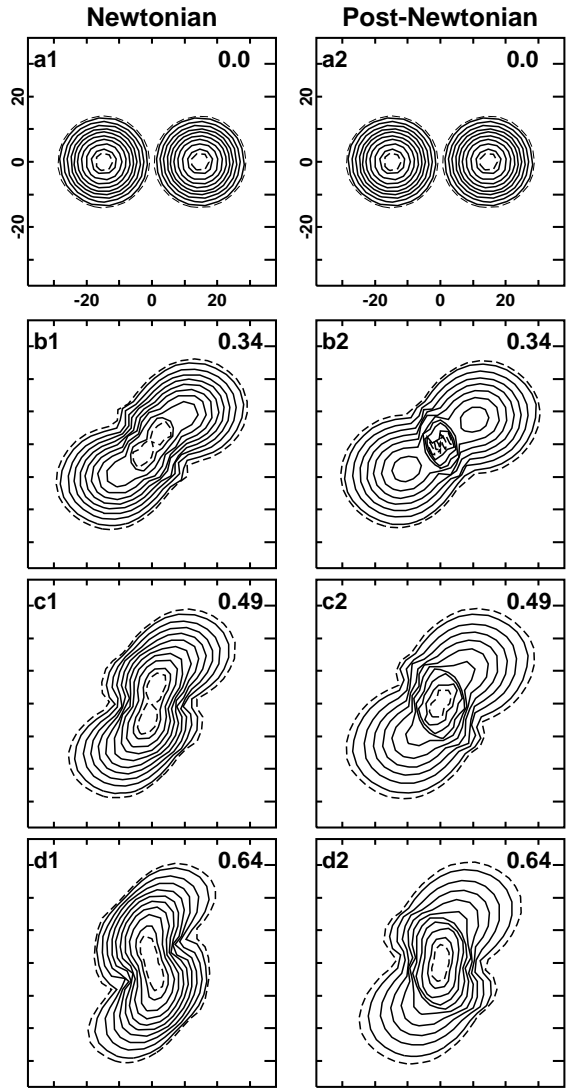}
    \caption{Density on the $x$-$y$ plane. The left and
      right figures are for the Newtonian(N) and post-Newtonian(PN)
      calculations, respectively. Notations are the same as for
      Fig.\protect\ref{fig:eqsp2-density}.}
    \label{fig:pndn-density}
  \end{center}
\end{figure}
\begin{figure}[p]
  \begin{center}
    \leavevmode
    \epsfysize=.9\textheight \addtolength{\epsfysize}{-4em}
    \epsfbox{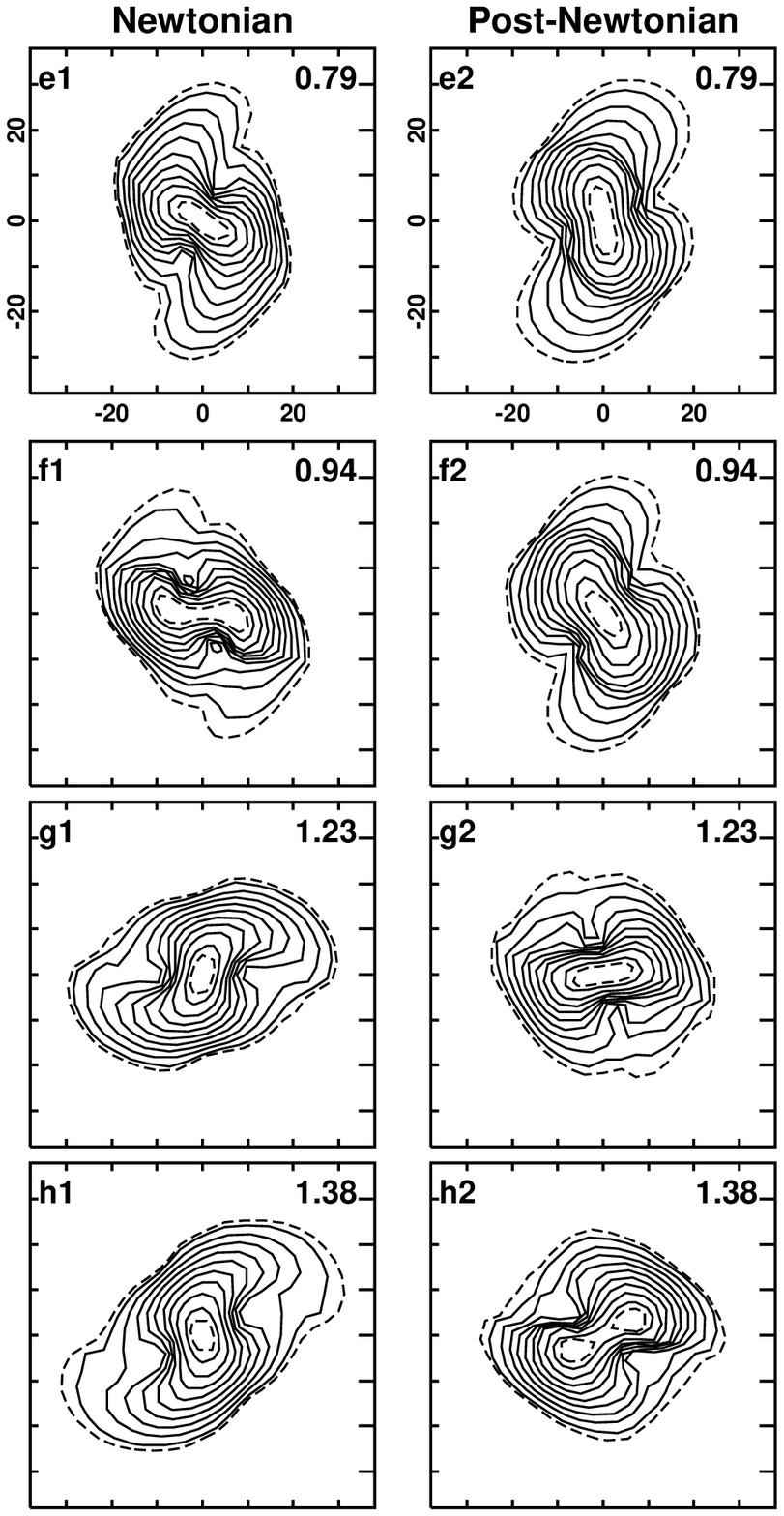}
  \end{center}
\end{figure}%
\begin{figure}[p]
  \begin{center}
    \leavevmode
    \epsfysize=.9\textheight \addtolength{\epsfysize}{-4em}
    \epsfbox{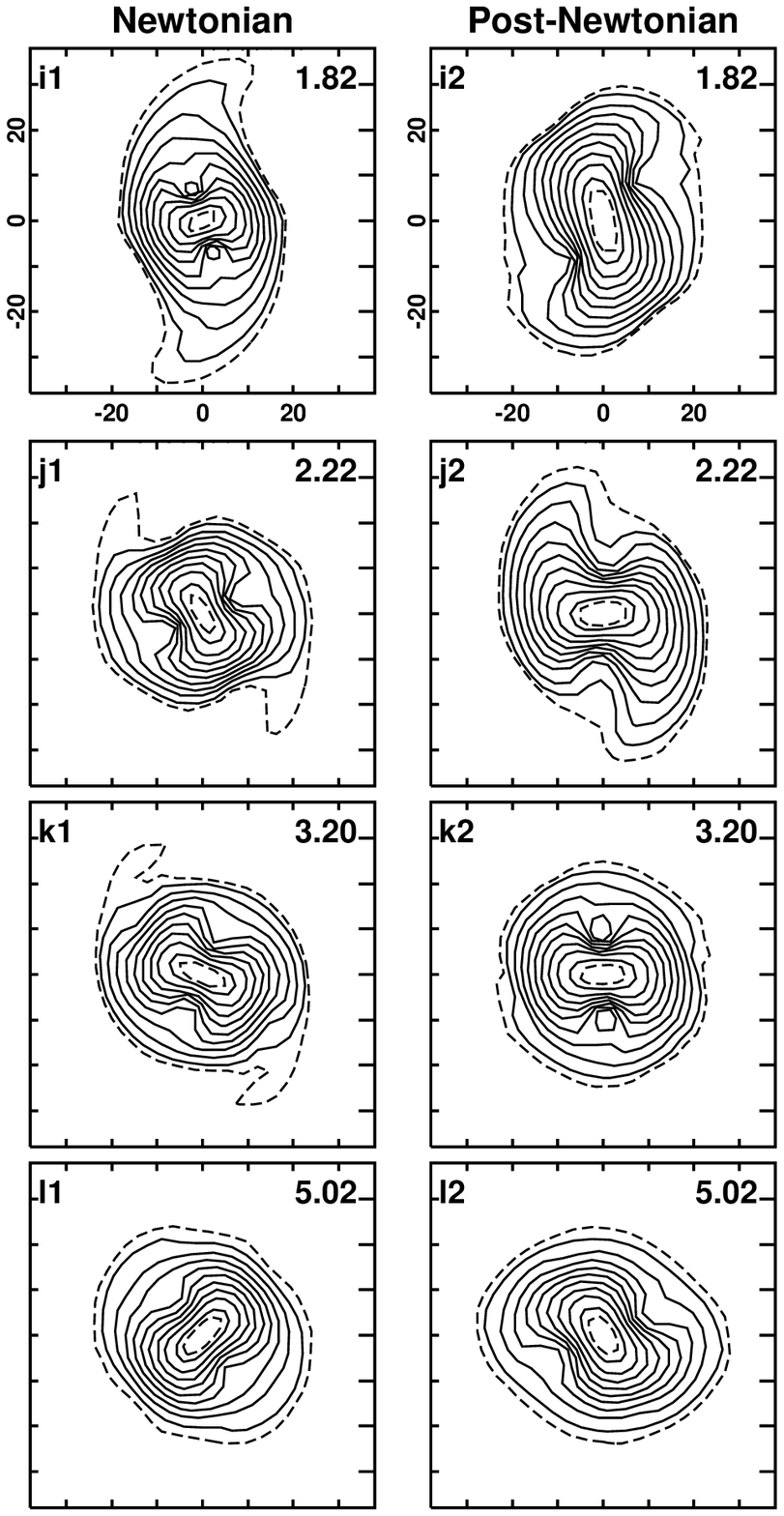}
  \end{center}
\end{figure}%
\begin{figure}[tbp]
  \begin{center}
    \leavevmode
    \epsfxsize=.8\textwidth \epsfbox{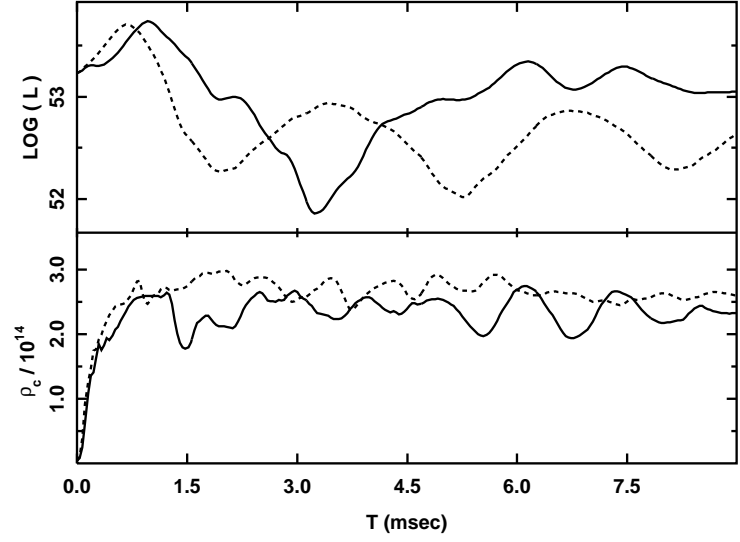}
    \caption{Luminosity (in units of erg/sec) and central density (in
      units of g/cm$^3$) as functions of time (in units of
      millisecond). The solid and dashed lines are for PN and N,
      respectively.} \label{fig:pndn-flux}
  \end{center}
\end{figure}
\begin{figure}[tbp]
  \begin{center}
    \leavevmode
    \epsfxsize=.8\textwidth\epsfbox{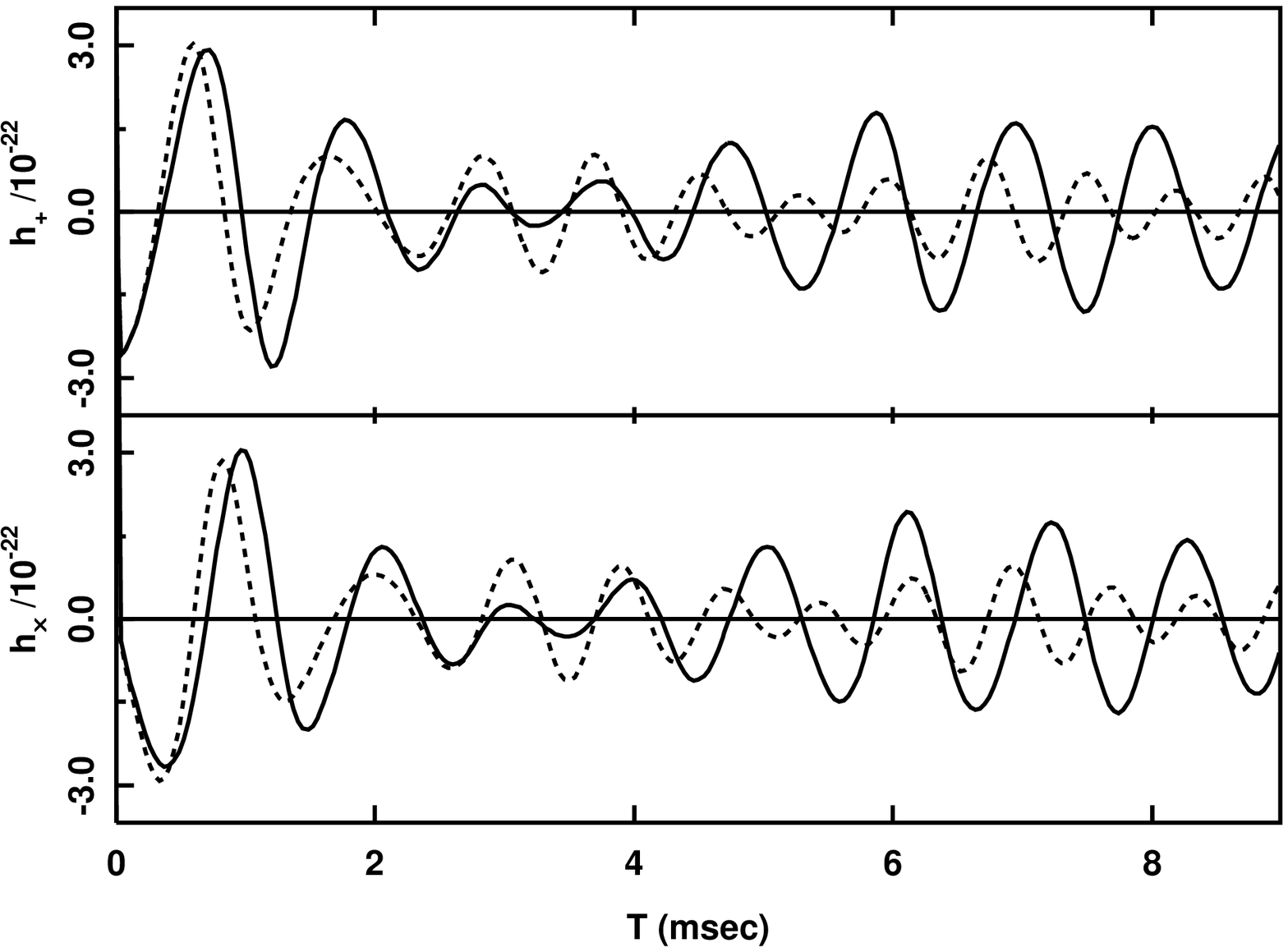}
    \caption{Wave forms of $h_{+}$ and $h_{\times}$ observed on the
      $z$-axis at 10Mpc. The solid and dashed lines are for PN and N,
      respectively.}  \label{fig:pndn-wavef}
  \end{center}
\end{figure}%

Figure \ref{fig:pndn-density} shows the evolution of the density and
the velocity on the $x$-$y$ plane and Fig.\ref{fig:pndn-flux} shows
the emitted energy $L$ and the central density $\rho_c$ as functions
of time. \Video{PN} The result of the post-Newtonian(PN) calculation
is shown on the right of Fig.\ref{fig:pndn-density} and as solid lines
in Fig.\ref{fig:pndn-flux}, while the result of the Newtonian(N)
calculation is on the left and as dashed lines. In PN, coalescence
begins more rapidly since general relativity effectively increases the
gravitational force. Then a strong shock appears which makes the
matter reexpand, and a radial oscillation with large amplitude is
excited. The coalescence-expansion oscillation lasts longer in the PN
case than in the N case. As a result the total energy emitted in the
gravitational radiation increases in comparison with the N case; $3.4
\times 10^{50}$erg in PN and $2.2 \times 10^{50}$erg in N up to
8.6msec.  Figure \ref{fig:pndn-wavef} shows the wave form observed
along the $z$-axis, where solid lines are for PN and dashed lines are
for N. Although $h$ is damped faster in N, the maximum amplitudes are
almost the same. The appearance of the strong shock in the PN
calculation has little effect on the wave form.

The difference between PN and N is caused principally by the strong
shock which appears in the initial stage. Thus if the initial data
consists of two stars in rotational equilibrium and coalescence
proceeds more slowly, the difference between PN and N may be less
pronounced. However we do not consider the red-shift effect and
excitation of the quasi-normal modes of the resulting black hole or
neutron star, which may be essential in the gravitational radiation
\cite{SP85,NOK87}. To complete the program of studying a final destiny
of a coalescing binary system and computing emission of gravitational
waves in such a event, we should solve a full set of the Einstein
equation.

\section{GENERAL RELATIVISTIC SIMULATIONS}

In this section, we discuss recent developments of our fully general
relativistic code for numerical calculation of {\em the last three
milliseconds} of coalescing binary neutron stars.

\subsection{Basic Equations}

\subsubsection{Initial Value Equations}

In the (3+1)-formalism of the Einstein equation, the line element
takes the form
\begin{equation}
  ds^2 = - \alpha^2 dt^2 + \gamma_{ij} ( dx^i + \beta^i dt )
  ( dx^j + \beta^j dt),
\end{equation}
where $\alpha$, $\beta^i$ and $\gamma_{ij}$ are the lapse function,
the shift vector and the intrinsic metric of 3-space, respectively.
The initial data should satisfy the constraint equations
\begin{eqnarray}
  & & R + K^2 - K_{ij} K^{ij} = 16 \pi \rhoh , \label{eq:hconst} \\
  & & D_j \left( K^j{}_i - \delta^j{}_i K \right) = 8 \pi J_i,
      \label{eq:mconst}
\end{eqnarray}
where $R$ is the 3-dimensional Ricci scalar curvature, $K_{ij}$ is the
extrinsic curvature, $K$ is its trace and the symbol $D_i$ denotes the
covariant derivative with respect to $\gamma_{ij}$.  The quantities
$\rhoh$ and $J_i$ are, respectively, the energy density and the
momentum density of the matter measured by the normal line observer.
The relation of these quantities to the stress-energy tensor is given
by Eq.(\ref{eq:qhydro}) below.

Now we assume that on the Cauchy slice the trace $K = 0$ and
$\gamma_{ij}$ is conformally flat, that is,
\begin{equation}
  \gamma_{ij} = \phi^4 \, \gammaT_{ij},
\end{equation}
where $\gammaT_{ij}$ is the flat metric. Defining the conformal
transformation as
\begin{equation}
  \label{eq:conf-i}
  \begin{array}[c]{l}
    \KT_{ij}  \equiv \phi^2     K_{ij}, \ \
    \KT_i{}^j \equiv \phi^6     K_i{}^j \ \
    \KT^{ij} \equiv  \phi^{10}  K^{ij} \\
    \rhob \equiv \phi^6  \rhoh \ \ \mbox{and} \ \ 
    \JT_i \equiv \phi^6  J_i ,
  \end{array}
\end{equation}
Eq.(\ref{eq:mconst}) can be expressed as
\begin{equation}
  \label{eq:mconst2}
  \DT_j \KT^j{}_i = 8 \pi \JT_i ,
\end{equation}
where $\DT_i$ is the covariant derivative with respect to
$\gammaT_{ij}$ \cite{York79}. The traceless extrinsic curvature can be
decomposed with the transverse traceless part
$\KT^{\rscript{TT}}_{ij}$ and the longitudinal traceless part
$(LW)_{ij}$ \cite{York73};
\begin{equation}
  \KT_{ij} = \KT^{\rscript{TT}}_{ij} + (LW)_{ij} ,
\end{equation}
where
\begin{equation}
  (LW)_{ij} = \DT_i W_j + \DT_j W_i
  - \frac{2}{3} \, \gammaT_{ij} \DT^{\ell} W_{\ell} .
\end{equation}
Assuming $\KT^{\rscript{TT}}_{ij} = 0$, Eq.(\ref{eq:mconst2}) is
reduced to
\begin{equation}
  \label{eq:mconst3}
  \LaplaceT W_i + \frac{1}{3} \DT_i \DT^j W_j =
  8 \pi \JT_i ,
\end{equation}
where $\widetilde{\triangle} \equiv \DT^i \DT_i$.  Equation
(\ref{eq:mconst3}) is the coupled elliptic equation but, by defining
\begin{equation}
  \label{eq:divW}
  \chi \equiv \DT_i W^i,
\end{equation}
it can be decoupled into
\begin{equation}
  \label{eq:DdivW}
  \LaplaceT \chi = 6 \pi \DT_i \JT^i
\end{equation}
and
\begin{equation}
  \label{eq:DW}
  \LaplaceT W_i = 8 \pi \JT_i - \frac{1}{3} \DT_i \chi .
\end{equation}
The boundary conditions for $\chi$ and $W_i$ are
\begin{equation}
  \chi = O \left( \frac{1}{r^4} \right)
\end{equation}
and
\begin{equation}
  W_i = \frac{\epsilon_{ijk} x^j M^k}{r^3}
  + O \left( \frac{1}{r^3} \right),
\end{equation}
where $M^k$ is a constant related to the angular momentum of the
system. 

The conformal transformation of the scalar curvature is
\begin{equation}
  R = \phi^{-4} \widetilde{R} - 8 \phi^{-5} \LaplaceT \phi,
\end{equation}
where $\widetilde{R}$ is the scalar curvature with respect to
$\gammaT_{ij}$. Since $\widetilde{R} = 0$ for a conformally flat
metric, Eq.(\ref{eq:hconst}) is reduced to
\begin{equation}
  \label{eq:hconst2}
  \LaplaceT \phi = -2 \pi \phi^{-1} \rhob - \frac{1}{8} \phi^{-7}
  \KT_{ij} \KT^{ij}. 
\end{equation}
The boundary condition for $\phi$ is
\begin{equation}
  \phi = 1 + \frac{M_G}{2r} + O \left( \frac{1}{r^3} \right) ,
\end{equation}
where $M_G$ is the gravitational mass of the system.

\subsubsection{Relativistic Hydrodynamics}

We assume the perfect fluid stress-energy tensor, which is given by
\begin{equation}
  \label{eq:enmom}
  T_{\mu \nu} = ( \rho + \rho \varepsilon + p ) u_{\mu} u_{\nu}
  + p g_{\mu \nu},
\end{equation}
where $\rho$, $\varepsilon$ and $p$ are the proper mass density, the
specific internal energy and the pressure, respectively, and $u_\mu$
is the four-velocity of the fluid. The energy density $\rhoh$, the
momentum density $J_i$ and the stress tensor $S_{ij}$ of the matter
measured by the normal line observer are, respectively, given by
\begin{equation}
  \label{eq:qhydro}
  \rhoh \equiv n^\mu n^\nu T_{\mu \nu}, \ \
  J_i \equiv - h_i{}^\mu n^\nu T_{\mu \nu} \ \ \mbox{and} \ \ 
  S_{ij} \equiv h_i{}^\mu h_j{}^\nu T_{\mu \nu},
\end{equation}
where $n_\mu$ is the unit timelike four-vector normal to the spacelike 
hypersurface and $h_{\mu \nu}$ is the projection tensor into the
hypersurface defined by
\begin{equation}
  h_{\mu \nu} = g_{\mu \nu} + n_\mu n_\nu .
\end{equation}
The relativistic hydrodynamics equations are obtained from the
conservation of baryon number, $\nabla_\mu ( \rho u^\mu )$, and the
energy-momentum conservation law, $\nabla_\nu T_\mu{}^\nu{}$. In order
to obtain equations similar to the Newtonian hydrodynamics equations,
we define $\rhon$ and $u_i^N$ as
\begin{equation}
  \rhon \equiv \sgamma \auz \rho \ \ \  \mbox{and} \ \ \  
  u_i^N = \frac{J_i}{\auz \rho},
\end{equation}
respectively, where $\gamma = \mbox{det} (\gamma_{ij})$. Then the
equation for the conservation of baryon number takes the form
\begin{equation}
  \label{eq:hydrob}
  \partial_t \rhon +   \partial_\ell \left( \rhon V^\ell \right) = 0 ,
\end{equation}
where
\begin{equation}
  V^\ell = \frac{u^\ell}{u^0} = \frac{\alpha J^\ell}{p + \rhoh} .
\end{equation}
The equation for momentum conservation is
{\arraycolsep = 2pt
\begin{eqnarray}
  \partial_t (\rhon u_i^N) +
  \partial_\ell \left( \rhon u_i^N V^\ell \right)
  & = & - \sgamma \alpha \partial_i p - \sgamma ( p + \rhoh )
  \partial_i \alpha \nonumber \\
  & & \frac{\sgamma \alpha J^k J^\ell}{2(p + \rhoh)} \partial_i
  \gamma_{k \ell} + \sgamma J_\ell \partial_i \beta^\ell .
  \label{eq:hydrom}
\end{eqnarray}
The equation for internal energy conservation becomes
\begin{equation}
  \label{eq:hydroe}
  \partial_t (\rhon \varepsilon) +
  \partial_\ell \left( \rhon \varepsilon V^\ell \right) =
  - p \frac{1}{\alpha \sgamma} \partial_\nu
  \left( \sgamma \alpha u^\nu \right).
\end{equation}
To complete hydrodynamics equations, we need an equation of state,
\begin{equation}
  \label{eq:eos}
  p = p(\varepsilon, \rho).
\end{equation}
Equations (\ref{eq:hydrob})--(\ref{eq:hydroe}), whose structure is the
same as Eqs.(\ref{eq:evol})--(\ref{eq:enrg}), can be solved using the
post-Newtonian hydrodynamics code described the previous section.

\subsubsection{Time Evolution of the Extrinsic Curvature}

The evolution equation for the extrinsic curvature takes the form
\begin{equation}
  \label{eq:evolk}
  \partial_t K_{ij} - \partial_\ell \left( K_{ij} \beta^\ell \right)
  = (SK)_{ij}^{\rscript{main}} + (SK)_{ij}^{\rscript{NL}}
  + (SK)_{ij}^{\beta},
\end{equation}
where
{\arraycolsep = 2pt
\begin{eqnarray}
  (SK)_{ij}^{\rscript{main}} & = & \alpha \left\{ R_{ij}
    - 8 \pi \left[ S_{ij} + {\textstyle \frac{1}{2}} \gamma_{ij}
      \left( \rhoh - S^\ell{}_\ell \right) \right] \right\}
  - D_i D_j \alpha \\[.25em]
  (SK)_{ij}^{\rscript{NL}} & = & \alpha \left( K K_{ij}
    - 2 K_{i \ell} K^\ell{}_j \right), \\[.25em]
  (SK)_{ij}^\beta & = & K_{mi} \partial_j \beta^m
  + K_{mj} \partial_i  \beta^m - K_{ij} \partial_m \beta^m,
\end{eqnarray}
and $R_{ij}$ is the 3-dimensional Ricci tensor. Equation
(\ref{eq:evolk}) can be solved in the same way as the hydrodynamics
equations with the velocity $- \beta^\ell$ instead of $V^\ell$.

\subsubsection{Time Slicing}

Defining the conformal factor $\phi$ as
\begin{equation}
  \phi \equiv \left( \mbox{det} \left( \gamma_{ij} \right)
  \right)^{\frac{1}{12}},
\end{equation}
we use the lapse function $\alpha$ given by
\begin{equation}
  \label{eq:confalpha}
  \alpha = \exp \left[ -2 \left( (\phi - 1) +
      \frac{(\phi - 1)^3}{3} + \frac{(\phi - 1)^5}{5} \right)
  \right] .
\end{equation}
Shibata and Nakamura\cite{SN92} called this time slicing as the
conformal time slicing because $\alpha$ is determined by the conformal
factor $\phi$. In the conformal time slicing, the space outside the
central matter quickly approaches the Schwarzschild metric. Detailed
discussion on the nature of the conformal time slicing is given in the
reference \cite{SN92}.  The equation for the conformal factor $\phi$
is obtained from the Hamiltonian constraint equations,
\begin{equation}
  \label{eq:conf-t}
  \LaplaceT \phi = - \frac{\phi^5}{8} \left( 16 \pi \rhoh
    + K_{ij} K^{ij} - K^2 - \phi^{-4} R \right),
\end{equation}
where $\LaplaceT$ is the Laplacian with respect to
\begin{equation}
  \label{eq:defgammat}
  \gammaT_{ij} = \phi^{-4} \gamma_{ij}.
\end{equation}
Note that $\gammaT_{ij}$ is {\em not} the flat-space metric for
$t \ne 0$.

\subsubsection{Spatial Coordinates}

Here we define conformal transformation, different from
Eq.(\ref{eq:conf-i}), as
\begin{equation}
  \label{eq:conformal}
  \begin{array}[c]{l}
    \KT_{ij}  \equiv \phi^{-4}     K_{ij}, \ \
    \KT_i{}^j \equiv K_i{}^j, \ \
    \KT \equiv  K, \\[.25em]
    \betaT_i \equiv \phi^{-4} \beta_i \ \ \mbox{and} \ \ 
    \betaT^i \equiv \beta^i ,
  \end{array}
\end{equation}
and the equation for metric thereby becomes
\begin{eqnarray}
  \label{eq:gammat}
  \partial_t \gammaT_{ij} & = & \left( \DT_i \betaT_j + \DT_j \betaT_i
    - \frac{2}{3} \gammaT_{ij} \DT_\ell \betaT^\ell \right)
  - 2 \alpha \left( \KT_{ij} - \frac{1}{3} \gammaT_{ij} \KT \right) \\
  & \equiv & A_{ij}^T, \nonumber
\end{eqnarray}
where $\DT_i$ denotes the covariant derivative with respect to
$\gammaT_{ij}$ defined by Eq.(\ref{eq:defgammat}).  Now we demand the
divergence with respect to the flat-space metric of the right hand
side of Eq.(\ref{eq:gammat}) to vanish,
\begin{equation}
  \label{eq:pmind}
  \partial_i A_{ij}^T = 0,
\end{equation}
which is reduced to the equation for the shift vector $\beta^i$,
\begin{eqnarray}
  \lefteqn{\nabla^2 \beta^i + \frac{1}{3} \partial_i
    \left( \partial_\ell \beta^\ell \right) = } \nonumber \\
  & & \partial_j \left[ 2 \alpha \left( \KT_{ij} - \frac{1}{3}
      \gammaT_{ij} K \right) \right] \label{eq:beta} \\
  & - & \partial_j \left[ \gammaH_{j \ell} \partial_i \beta^\ell
    + \gammaH_{i \ell} \partial_j \beta^\ell
    - \frac{2}{3} \gammaH_{ij} \partial_\ell \beta^\ell
    + \beta^\ell \partial_\ell \gammaH_{ij} \right], \nonumber
\end{eqnarray}
where $\nabla^2$ is the simple flat-space Laplacian and $\gammaH_{ij}
= \gammaT_{ij} - \delta_{ij}$.  Equation (\ref{eq:beta}) is the
similar to Eq.(\ref{eq:mconst3}) while $\beta^i$ appears in the right
hand side, too.

If we demand
\begin{equation}
  \label{eq:mind}
  \DT^j A^T_{ij} = 0,
\end{equation}
instead of Eq.(\ref{eq:pmind}), we have the minimal distortion
condition proposed by York and Smarr \cite{YS78}. In this sense, we
call Eq.(\ref{eq:pmind}) the pseudo-minimal distortion condition.

\subsection{Numerical Methods}

We use Cartesian $(x,y,z)$ isotropic coordinates for relativistic
simulation, too. The hydrodynamics equations,
Eqs.(\ref{eq:hydrob})--(\ref{eq:hydroe}), and the evolution equations
for the extrinsic curvature, Eq.(\ref{eq:evolk}), and the 3-metric,
Eq.(\ref{eq:gammat}), are solved through the same scheme as in the
Newtonian simulation. The elliptic equations are solved through the
MICCG method. The equation for the conformal factor in the initial
hypersurface, Eq.(\ref{eq:hconst2}), can be reduced to
\begin{equation}
  \label{eq:hconst3}
  \nabla^2 \phi = 4 \pi \rho_1(\phi),
\end{equation}
since the initial hypersurface is conformally flat ($\LaplaceT =
\nabla^2$). The source term $\rho_1(\phi)$ is a non-linear function of
$\phi$. This equation is solved by an iterative method: The following
iteration is repeated until convergence,
\begin{equation}
  \phi^{(I+1)} = (\nabla^2)^{-1} \left[ 4 \pi \rho_1
    \left( \phi^{(I)} \right)   \right]
  \ \ \  ( \mbox{for} \  I = 1, 2, \cdots ).
\end{equation}
Equations (\ref{eq:conf-t}) and (\ref{eq:beta}) for $t > 0$ are
also non-linear or coupled elliptic equations, which can be written as
\begin{equation}
  \label{eq:conformal2}
  \nabla^2 \phi = 4 \pi \rho_2 (\phi),
\end{equation}
\begin{equation}
  \label{eq:beta2-1}
  \nabla^2 \chi = 3 \pi \partial_i \left( \rho_3^i (\beta) \right)
\end{equation}
and
\begin{equation}
  \label{eq:beta2-2}
  \nabla^2 \beta^i = - \frac{1}{3} \partial_i \chi
  + 4 \pi \rho_3^i (\beta),
\end{equation}
where $\chi \equiv \partial_\ell \beta^\ell$.  The cost of finding
self-consistent solutions for
Eqs.(\ref{eq:conformal2})--(\ref{eq:beta2-2}) at each time step is
very expensive and hence we use the values at the previous time step
for $\phi$ and $\beta^i$ in the source terms $\rho_2$ and $\rho_3$.
The accuracy of solutions by this method seems to be sufficient to
examine the global behavior of the numerical schemes and the
coordinate conditions. The method should be improved in a future
version.

Although any equation of state can be used, for the present test
calculation, we use a simple equations state Eq.(\ref{eq:eosg2}) with
$\gamma = 2$ to express a hard equation of state for a neutron star.

\subsection{Numerical Results}

Now we show results of test calculations. Here we use units of
\begin{equation}
  M = \sol, \ \ L = \frac{G \sol}{c^2} \ \ \mbox{and} \ \ 
  T = \frac{G \sol}{c^3}.
\end{equation}
In the following test calculations, we use a $80 \times 80 \times 80$
grid covering $[-40, 40]$ in each direction.

\subsubsection{Spherically Symmetric Dust Collapse}

In the conformal time slicing given by Eq.(\ref{eq:confalpha}), the
time-variable parts of $\gammaT_{ij}$ can be considered as
gravitational radiation parts, since the space outside the central
matter approaches the Schwarzschild metric quickly \cite{SN92}. Thus
the total energy of the gravitational waves is given by
\begin{equation}
  \label{eq:gwtotal}
  E_{\rscript{GW}} = \int dt \int_{r \rightarrow \infty}
    d \Omega \, \frac{r^2}{32\pi} \left( A^{\rscript{TT}}_{ij} \right)^2,
\end{equation}
where $A^{\rscript{TT}}_{ij}$ is the transverse-traceless part of the time
derivative of $\gammaT_{ij}$,
\begin{equation}
  A^{\rscript{TT}}_{ij} = \left( \partial_t \gammaT_{ij}
  \right)^{\rscript{TT}}
\end{equation}
and the time derivative of $\gammaT_{ij}$ is given by
Eq.(\ref{eq:gammat}).  Now we define the ``energy density of the
gravitational waves'', although it cannot be defined locally, as
\begin{equation}
  \label{eq:gwdens}
  \rho_{_{\rscript{GW}}} = \frac{1}{32\pi} \left
    ( A^{\rscript{TT}}_{ij} \right)^2 
  = \frac{1}{32\pi} (A^{\rscript{TT}})_{ij} (A^{\rscript{TT}})^{ij}.
\end{equation}

\begin{figure}[tbp]
  \begin{center}
    \leavevmode
    \epsfxsize=\textwidth \epsfbox{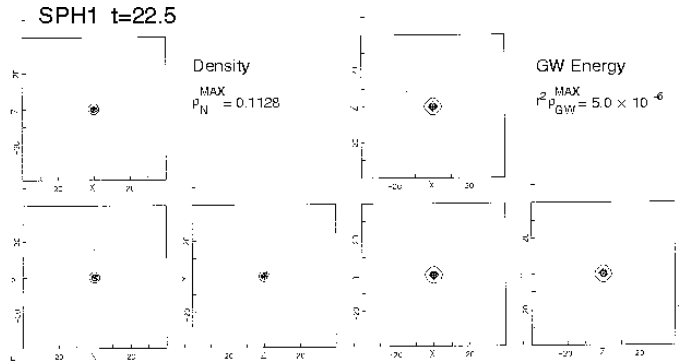}
    \caption{Mass density $\rhon$ (left) and ``energy density of the
      gravitational waves'' $r^2 \rho_{_{GW}}$ (right) on the
      $x$-$y$, $y$-$z$ and $x$-$z$ planes for a spherically symmetric
      dust collapse. The time in units of $\sol$, and the maximum
      value of density in units of $\sol^{-2}$ are shown.}
    \label{fig:sdust}
  \end{center}
\end{figure}

In order to examine whether $A_{ij}^{\rscript{TT}}$ shows the
gravitational waves' degree of freedom, we performed calculation of
collapse of a spherically symmetric dust star of mass $M$. Figure
\ref{fig:sdust} shows the mass density $\rhon$ and the ``energy
density of the gravitational waves'' $ r^2 \rho_{_{\rscript{GW}}}$.
Although $r^2 \rho_{_{\rscript{GW}}}$ has a non-zero value near the
center, it does not propagate outward. Since the meaning of
``transverse-traceless'' is clear only in the asymptotically flat
region, the results is consistent with the idea that
$A_{ij}^{\rscript{TT}}$ represents the gravitational waves. The total
energy of the gravitational waves $E_{\rscript{GW}}$ is $\sim 10^{-7}
M$ at $t = 30M$ observed at $r = 20M$. This value is essentially zero
within numerical errors.

\subsubsection{Formation of a Rotating Black Hole}

Stark and Piran \cite{SP85} performed axially symmetric simulations
for formation of a black hole and directly computed their
gravitational radiation emission. For comparison with their results,
we set similar initial conditions, although our coordinate condition
is different from theirs. We use a simple equation of state,
Eq.(\ref{eq:eosg2}), which is the same as Stark and Piran. We place a
spherically symmetric star of $\gamma = 2$ polytrope of mass $1\sol$
with internal energy necessary for Newtonian equilibrium. The rotation
velocity with constant $\Omega$ is added to this system. We performed
four simulations, ROT1, ROT2, ROT3 and ROT4, each of which is
characterize by the initial value of $a/M_g$, where $a$ is the angular
momentum per unit gravitational mass $M_g$. The value of $a/M_g$ is
0.3, 0.5, 0.8 and 1.0, respectively.

\begin{figure}[tbp]
  \begin{center}
    \leavevmode
    \epsfxsize=\textwidth \epsfbox{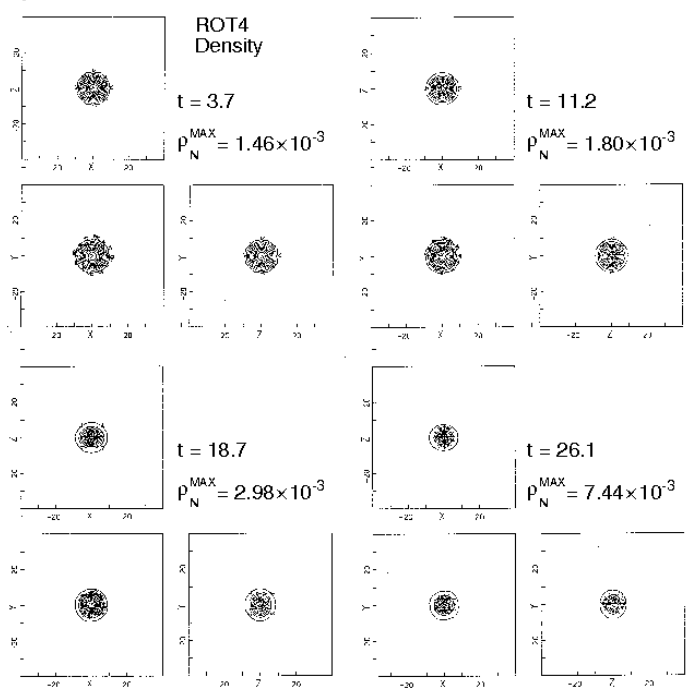}
    \caption{Mass density $\rhon$ on the $x$-$y$, $y$-$z$ and $x$-$z$
      planes for a spherically symmetric collapse of a rotating star
      ROT4. Arrows indicate the velocity vector $V^i$.}
    \label{fig:rot4-dv}
  \end{center}
\end{figure}
\begin{figure}[tbp]
  \begin{center}
    \leavevmode
    \epsfxsize=\textwidth \epsfbox{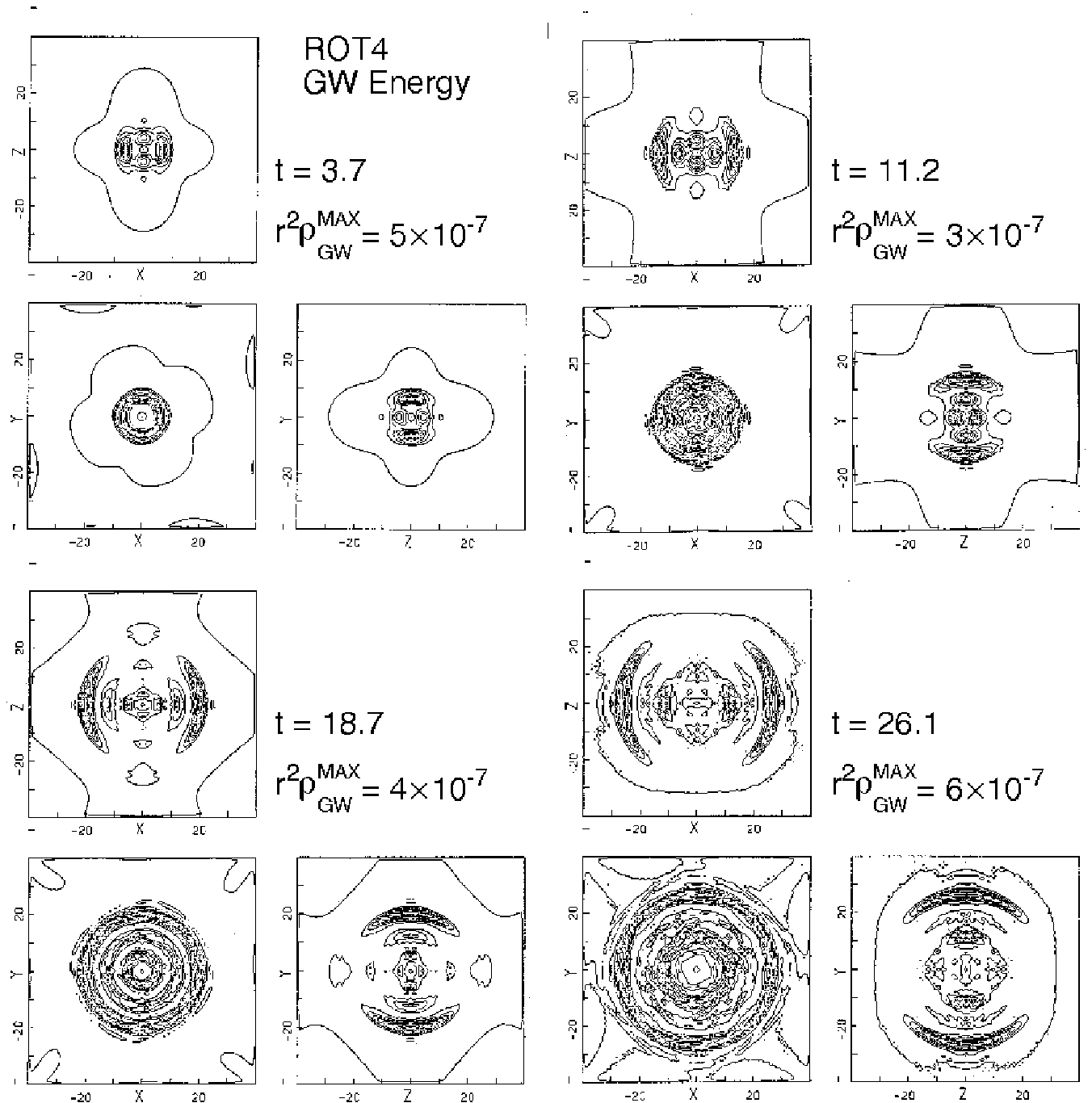}
    \caption{``Energy density of the gravitational waves'' $r^2
      \rho_{_{GW}}$ for ROT4.}
    \label{fig:rot4-gwe}
  \end{center}
\end{figure}

Figure \ref{fig:rot4-dv} show the evolution of density distribution of
ROT4. It represents that the centrifugal force prevents the star from
collapse in the direction perpendicular to the rotation axis and the
star becomes an oblate spheroid. The ``energy of gravitational waves''
is shown in Fig.\ref{fig:rot4-gwe}. A pattern almost axially symmetric
with a few peaks appears on the $x$-$y$ plane. Deviation from axial
symmetry is caused by coarseness of our Cartesian grid.  The wave
pattern on $x$-$z$ and $y$-$z$ planes looks like the quadrupole
emission proportional to $\sin^4 \theta$, which is also the case in
Stark and Piran. The total energy radiated at $t = 30 \sol$ observed
at $r = 20\sol$ are $5.7 \times 10^{-7} \sol$, $3.8 \times 10^{-6}
\sol$, $1.5 \times 10^{-5} \sol$ and $3.6 \times 10^{-5} \sol$ for
ROT1, ROT2, ROT3 and ROT4, respectively. Stark and Piran found that
the total energy of gravitational waves is proportional to $a^4$ for
small $a$ and it levels off at $a \sim M_g$. The same dependency
reappears in our calculation; the total energy is expressed
approximately
\begin{equation}
  E_{\rscript{GW}} = 5 \times 10^{-5} \left( \frac{a}{M_g} \right)^4 .
\end{equation}
We have to remark that the value of the coefficient in Stark and Piran
is about 20 times larger than ours. However, this is not disagreement,
because Stark and Piran pursued their calculation up to $t = 100 M$
while we did only up to $t = 30M$. The energy we observed is thus just
expected from the time dependence of the gravitational wave energy
given by Stark and Piran \cite{SP85}.

\subsubsection{Coalescing Binary Neutron Stars}

Now we begin to attack the problem of coalescing binary neutron stars.
We place as initial data two spherical neutron stars of mass $M =
1.0\sol$ and radius $r_0 = 6M$ with density distribution ($\rhon$) of
$\gamma = 2$ polytrope at $x = r_0$ and $x = - r_0$; two stars just
touch each other. We add a rigid rotation with angular velocity
$\Omega$ as well as an approaching velocity $v_a$ to this system such
that
\begin{eqnarray}
  v_x^N & = & \left\{
    \begin{array}[c]{ll}
      - \Omega y - v_a \ & \mbox{if} \ x > 0, \\[.5em]
      - \Omega y + v_a \ & \mbox{if} \ x < 0,
    \end{array} \right. \\[.5em]
  v_y^N & = & \Omega x,
\end{eqnarray}
setting $v_a = \Omega r_0$.

We performed three simulations, BI1, BI2 and BI3, with different
values of $\Omega$. The total angular momentum divided by the square
of the gravitational mass are $0.71$, $1.43$ and $1.01$ for BI1, BI2
and BI3, respectively.

Figure \ref{fig:bi3-dens} shows the evolution of the density on the
$x$-$y$, $y$-$z$ and $x$-$z$ planes of BI3. The final object will be a
rotating black hole although we have not yet tried to determine the
apparent horizon. The ``gravitational wave energy'' $\rho_{_{GW}}$
shows an interesting feature of generation and propagation of the
waves. A spiral pattern appears on the $x$-$y$ plane while different
patterns with peaks around $z$-axis appear on the $x$-$z$ and $y$-$z$
planes. This can be explained naively by the quadrupole wave pattern
given by
\begin{equation}
  r^2 \rho_{_{GW}} = \frac{r^2}{32 \pi} \left( A^{\rscript{TT}}_{ij}
  \right)^2   \propto \cos^2 \theta + \sin^2 \theta 
  \sin^2 (2 \Omega(t - r) - 2   \varphi )/4.
\end{equation}
On the $x$-$y$ plane, where $\theta = \pi/2$, $\rho_{_{GW}}$ is
constant along the spiral of $r + \varphi/\Omega =$ constant, while
near $z$-axis, where $\theta \approx 0$, $\rho_{_{GW}} \propto \cos^2
\theta$.

At the numerical boundary, we have to impose the outgoing wave condition to
$\gammaT_{ij}$ and $K_{ij}$. In the present calculation, however,
$x_{\rscript{max}}$, $y_{\rscript{max}}$ and $z_{\rscript{max}}$ are
comparable to the wave length of the expected gravitational waves,
$\sim 40\sol$. We tried various boundary conditions, but none works
well. We therefore adopt a simple extrapolation at the numerical
boundary and suspend calculations when the gravitational waves reach
the numerical boundary. We hope that the problem will be overcome if
the numerical boundary is located farther, where the outgoing wave
condition becomes appropriate.

The total energy of the gravitational radiation amounts to $3 \times
10^{-3} \sol$ at $t = 35 \sol$ for BI3.

\ack{This work is supported by the Grant-in-Aid for Scientific
  Research on Priority Area of Ministry Education (04234104).}

\begin{figure}[htbp]
  \begin{center}
    \leavevmode
    \epsfxsize=\textwidth \epsfbox{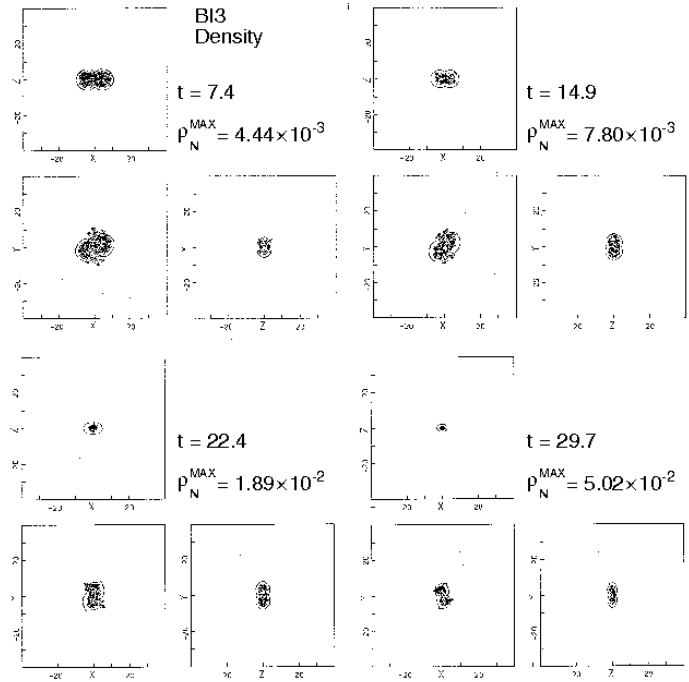}
    \caption{Density on the $x$-$y$, $y$-$z$ and $x$-$z$ planes for
      BI3. Arrows indicate the velocity vector $V^i$. }
    \label{fig:bi3-dens}
  \end{center}
\end{figure}
\begin{figure}[htbp]
  \begin{center}
    \leavevmode
    \epsfxsize=\textwidth \epsfbox{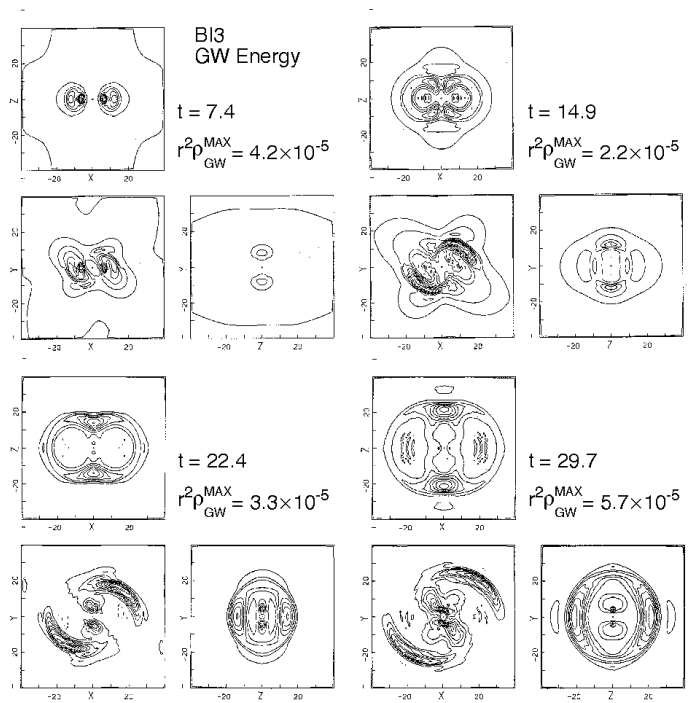}
    \caption{Energy density of the gravitational waves $r^2
      \rho_{_{GW}}$ on the $x$-$y$, $y$-$z$ and $x$-$z$ planes for
      BI3.}
    \label{fig:bi3-gwenergy}
  \end{center}
\end{figure}

\begin{figure}[htbp]

\end{figure}

\end{document}